\documentclass[journal]{IEEEtran}

\IEEEoverridecommandlockouts
\usepackage{cite}
\usepackage[dvipdf]{graphicx}
\usepackage[cmex10]{amsmath}
\usepackage{algorithmic}
\usepackage{algorithm}
\usepackage{array}
\usepackage{mdwmath}
\usepackage{mdwtab}
\usepackage[caption=false]{caption}
\usepackage{fixltx2e}
\usepackage{stfloats}
\ifCLASSOPTIONcaptionsoff
 \usepackage[nomarkers]{endfloat}
 \let\MYoriglatexcaption\caption
 \renewcommand{\caption}[2][\relax]{\MYoriglatexcaption[#2]{#2}}
\fi
\usepackage{url}

\newcommand{\defeq}{{\buildrel\Delta\over{=}}}

\begin{document}
%
\title{Doppler Spread Estimation by Tracking the Delay-Subspace for OFDM Systems in Doubly Selective Fading Channels}

\author{Xiaochuan~Zhao,~\IEEEmembership{Student Member,~IEEE,}
        Tao~Peng,~\IEEEmembership{Member,~IEEE,}
        Ming~Yang
        and~Wenbo~Wang,~\IEEEmembership{Member,~IEEE}
\thanks{Xiaochuan Zhao, Tao Peng, Ming Yang, and Wenbo Wang are with the Wireless Signal Processing and Network
Lab, and the Key Laboratory of Universal Wireless Communication,
Ministry of Education, Beijing University of Posts and Telecommunications, Beijing, China (e-mail: zhaoxiaochuan@gmail.com).}
\thanks{This work is sponsored in
part by the National Natural Science Foundation of China under grant
No.60572120 and 60602058, and by the national high technology
researching and developing program of China (National 863 Program)
under grant No.2006AA01Z257, and by the National Basic Research
Program of China (National 973 Program) under grant No.2007CB310602.
Part of the results in this paper was presented at IEEE Global
Communications Conference 2008 (IEEE GLOBECOM 2008), New Orleans,
LA, USA.}}
\maketitle
\begin{abstract}
In this paper, a novel maximum Doppler spread estimation algorithm
is presented for OFDM systems with the comb-type pilot pattern in
doubly selective fading channels. First, the least-squared estimated
channel frequency responses on pilot tones are used to generate two
auto-correlation matrices with different lags. Then, according to
these two matrices, a Doppler dependent parameter is measured. Based
on a time-varying multipath channel model, the parameter is expanded
and then transformed into a non-linear high-order polynomial
equation, from which the maximum Doppler spread is readily solved by
using the Newton's method. The delay-subspace is utilized to reduce
the noise that biases the estimator. Besides, the subspace tracking
algorithm is adopted as well to automatically update the
delay-subspace. Simulation results demonstrate that the proposed
algorithm converges for a wide range of SNR's and Doppler's.
\end{abstract}


\IEEEpeerreviewmaketitle

\section{Introduction}
The maximum Doppler spread is one of the key parameters for the
adaptive strategies to tune mobile communication systems to
accommodate various radio transmission environments \cite{Chu00}
and, especially, alleviate the inter-carrier interference (ICI) for
orthogonal frequency division multiplexing (OFDM) systems. Most
methods of estimating the maximum Doppler spread reported in
literatures are categorized into two classes \cite{Tepe01}:
level-crossing-rate-based and covariance-based techniques. Since the
algorithms reviewed in \cite{Tepe01}, such as \cite{Tepe01VT}, were
not specifically designed for OFDM systems, they did not exploit the
special signal structure of OFDM systems. On the other hand, most
algorithms designed for OFDM systems are covariance-based.
\cite{Scho02} determined the maximum Doppler spread through
estimating the smallest positive zero crossing point. Cai
\cite{Cai03} proposed to obtain the time correlation function (TCF)
by exploiting the cyclic prefix (CP) and its counterpart. However,
Yucek \cite{Yuce05} commented that for scalable OFDM systems whose
CP sizes were varying along the time, it was difficult for
\cite{Cai03} to sufficiently estimate the TCF, therefore its
accuracy would be degraded significantly. In stead, Yucek proposed
to estimate the maximum Doppler spread by taking advantage of the
periodic training symbols. However, in order to reduce overheads,
training symbols are sparse and typically transmitted as preambles
to facilitate the frame timing and carrier frequency
synchronization, which would cause \cite{Yuce05} to converge slowly
or even fail.

In this paper, we propose to estimate the maximum Doppler spread by
exploiting the comb-type pilot tones \cite{Cole02} that are widely
adopted by wireless standards. The channel frequency responses
(CFR's) estimated from the pilot tones are projected onto the
delay-subspace \cite{Sime04} to reduce the noise perturbation and
then used to acquire a Doppler dependent parameter. With a careful
expansion of the parameter, a nonlinear high-order polynomial
equation is formed, from which the maximum Doppler spread is readily
solved by resorting to the Newton's iteration. Moreover, the
subspace tracking algorithm \cite{Stro96} is adopted as well to
track the drifting delay-subspace.

This paper is organized as follows. In Section \ref{sec:model}, the
OFDM system and channel model are introduced. Then, the maximum
Doppler spread estimation algorithm is presented in Section
\ref{sec:dopplerest}. Simulation results and analyses are provided
in Section \ref{sec:simulations}. Finally, Section
\ref{sec:conclusion} concludes the paper.

\subsection{Basic Notation}
Uppercase and lowercase boldface letters denote matrices and column
vectors, respectively. $(\cdot)^H$ and $||\cdot||_F$ denote
conjugate transposition and Frobenius norm, respectively. $E(\cdot)$
represents the mathematical expectation of a stochastic process.
$[{\bf{\cdot}}]_{i}$ and $[{\bf{\cdot}}]_{i,j}$ denote the $i$-th
and $(i,j)$-th elements of a vector and a matrix, respectively.
$diag({\bf{A}})$ denotes a diagonal matrix with the diagonal
elements of ${\bf{A}}$ on the diagonal.

\setlength{\arraycolsep}{0.2em}

\section{System Model}
\label{sec:model} Consider an OFDM system with a bandwidth of
$BW=1/T$ Hz ($T$ is the sampling period). $N$ denotes the total
number of tones, and a CP of length $L_{cp}$ is inserted before each
symbol to eliminate the inter-block interference. Thus, the whole
symbol duration is $T_s=(1+r_{cp})NT$, where
$r_{cp}=\frac{L_{cp}}{N}$. In each OFDM symbol, $P$ tones are
transmitted as pilots to assist channel estimation. In addition, the
optimal pilot pattern, i.e., the equipowered and equispaced
\cite{Ohno04}, is assumed.

The complex baseband model of a linear time-variant mobile channel
with $L$ paths can be described by \cite{Stee92}
\begin{equation}
\label{eqn:channel}
{h(t,\tau)=\sum\nolimits_{l=0}^{L-1}{h_l(t)\delta\left({\tau-\tau_lT}\right)}}
\end{equation}
where $\tau_l \in {\mathcal{R}}$ is the non-sample-spaced delay of
the $l$-th path normalized to the sampling period, and $h_l(t)$ is
the corresponding complex amplitude. According to the assumption of
the wide-sense stationary uncorrelated scattering, $h_l(t)$'s are
modeled as uncorrelated narrowband complex Gaussian processes. In
the sequel, $P\geq{L}$ is assumed for determinability. Furthermore,
by assuming the uniform scattering environment introduced by Clarke
\cite{Clar68}, all $h_l(t)$'s have the identical normalized TCF,
therefore the TCF of the $l$-th path is
\begin{equation}
\label{eqn:rtdef}
{r_{t,l}({\Delta}t)=\sigma_l^2J_0\left(2{\pi}{f_d}{\Delta}t\right)}
\end{equation}
where $\sigma _l^2$ is the power of the $l$-th path, $f_d$ is the
maximum Doppler spread, and $J_0(\cdot)$ is the zeroth order Bessel
function of the first kind. Moreover, the total power of the channel
is normalized, i.e., $\sum\nolimits_{l=0}^{L-1}\sigma_l^2=1$.

Assuming a sufficient CP, i.e., $L_{cp} \geq L$, the discrete signal
model in the frequency domain is
\begin{equation}
\label{eqn:YmMatrixdef}
{\bf{y}}_f(n)={\bf{H}}_f(n){\bf{x}}_f(n)+{\bf{n}}_f(n)
\end{equation}
where
${\bf{x}}_f(n),{\bf{y}}_f(n),{\bf{n}}_f(n)\in{\mathcal{C}}^{N\times1}$
are the $n$-th transmitted and received signal and additive white
Gaussian noise (AWGN) vectors, respectively, and
${\bf{H}}_f(n)\in{\mathcal{C}}^{N\times{N}}$ is the channel transfer
matrix whose $(k+\nu,k)$-th element is
\begin{equation}
\label{eqn:Hdef}
\left[{\bf{H}}_f(n)\right]_{k+\upsilon,k}=\frac{1}{N}\sum\limits_{m=0}^{N-1}\sum\limits_{l=0}^{L-1}h_l(n,m)e^{-j2{\pi}({\upsilon}m+k{\tau_l})/N}\nonumber
\end{equation}
where $h_l(n,m)=h_l(nT_s+(L_{cp}+m)T)$ is the sampled complex
amplitude of the $l$-th path, and $k$ and $\upsilon$ denote
frequency and Doppler spread, respectively. Apparently, ICI is
present due to the non-diagonal ${\bf{H}}_f(n)$. However, when
$f_dT_s\le0.1$, the signal-to-interference ratio (SIR) is over 17.8
dB \cite{Choi01}, which enables us to discard non-diagonal elements
of ${\bf{H}}_f(n)$ with a negligible performance penalty.

For the comb-type pilot pattern, only pilot tones, denoted as
${\bf{y}}_{p}(n)\in\mathcal{C}^{P\times1}$, are extracted from
${\bf{y}}_f(n)$. Then, approximating ${\bf{H}}_f(n)$ by a diagonal
matrix, (\ref{eqn:YmMatrixdef}) is modified to
\begin{equation}
\label{eqn:Ypmdef}
{\bf{y}}_{p}(n)={\bf{X}}_{p}(n){\bf{h}}_{p}(n)+{\bf{n}}_{p}(n)
\end{equation}
where ${\bf{X}}_{p}(n)\in\mathcal{C}^{P\times{P}}$ is a pre-known
diagonal matrix, and ${\bf{h}}_{p}(n)\in\mathcal{C}^{P\times1}$
consists of the diagonal elements of ${\bf{H}}_f(n)$. By denoting
the instantaneous CFR as
$H(n,m,k)=\sum\nolimits_{l=0}^{L-1}h_l(n,m)e^{-j2\pi{k\tau_l}/N}$,
we have
$[{\bf{h}}_{p}(n)]_{p}=\frac{1}{N}\sum\nolimits_{m=0}^{N-1}H(n,m,\theta_p)$,
where $\theta_p$ is the index of the $p$-th pilot tone. Hence,
${\bf{h}}_{p}(n)$ is the time-averaging CFR during the $n$-th OFDM
symbol. Besides, the noise term
${\bf{n}}_p(n)\sim{\mathcal{CN}}(0,\sigma_n^2{\bf{I}}_P)$.

\section{Maximum Doppler Spread Estimation}
\label{sec:dopplerest} First of all, the least-squared (LS) channel
estimation on pilot tones is carried out at the receiver, that is,
\begin{equation}
\label{eqn:Hpmestdef}
{\bf{h}}_{p,ls}(n)={\bf{X}}_{p}(n)^{-1}{\bf{y}}_{p}(n)={\bf{h}}_{p}(n)+{\bf{w}}_{p}(n)
\end{equation}
where ${\bf{h}}_{p,ls}(n)\in\mathcal{C}^{P\times1}$ is the
LS-estimated time-averaging CFR, and
${{\bf{w}}_{p}(n)={\bf{X}}_{p}(n)^{-1}{\bf{n}}_{p}(n)}$ is the noise
term. For equipowered and PSK-modulated pilot tones,
${\bf{X}}_{p}(n)^H{\bf{X}}_{p}(n)={\bf{I}}_P$, therefore
${\bf{w}}_{p}(n)\sim\mathcal{CN}(0,\sigma_n^2{\bf{I}}_P)$.

By defining the Fourier transform matrix
${\bf{F}}_{p,\tau}\in\mathcal{C}^{P\times{L}}$ as
$[{\bf{F}}_{p,\tau}]_{p,l}=e^{-j2\pi\theta_p\tau_l/N}$, the 0-lag
auto-correlation matrix is
\begin{equation}
\label{eqn:Rh0def}
{\bf{R}}_{h}(0)\;{\defeq}\;E\left[{\bf{h}}_{p,ls}(n){\bf{h}}_{p,ls}(n)^H\right]=\xi_0{\bf{F}}_{p,\tau}{\bf{D}}{\bf{F}}_{p,\tau}^H+\sigma_n^2{\bf{I}}_P
\end{equation}
where ${\bf{D}}\;{\defeq}\;diag(\sigma_l^2)$, $l=0,\ldots,L-1$, and
$\xi_0$ is
\begin{equation}
\label{eqn:xi0def}
\xi_0\;\defeq\;\frac{1}{N^2}\sum\limits_{m=0}^{N-1}\sum\limits_{q=0}^{N-1}J_0(2{\pi}f_d(m-q)T)
\end{equation}

Similarly, the $\beta$-lag auto-correlation matrix of
${\bf{h}}_{p,ls}(n)$ is
\begin{equation}
\label{eqn:Rhbetadef}
{\bf{R}}_{h}(\beta)\;{\defeq}\;E\left[{\bf{h}}_{p,ls}(n+\beta){\bf{h}}_{p,ls}(n)^H\right]=\xi_{\beta}{\bf{F}}_{p,\tau}{\bf{D}}{\bf{F}}_{p,\tau}^H
\end{equation}
where $\xi_{\beta}$ is
\begin{equation}
\label{eqn:xibetadef}
\xi_{\beta}\;{\defeq}\;\frac{1}{N^2}\sum\limits_{m=0}^{N-1}\sum\limits_{q=0}^{N-1}J_0(2{\pi}f_d(m-q+\beta(1+r_{cp})N)T)
\end{equation}

If the number of channel taps is known, the delay-subspace, denoted
as ${\bf{U}}_{\tau,s}\in{\mathcal{C}}^{P\times{L}}$, and the
variance of noise $\sigma_n^2$ can be acquired with the eigenvalue
decomposition. If unknown, however, the number of significant taps
can be estimated by applying the minimum description length (MDL)
algorithm \cite{Wax85}. Then, a Doppler dependent parameter is
defined as
\begin{equation}
\label{eqn:etadef}
\eta\;\defeq{}\;\sqrt{\frac{||{\bf{U}}_{\tau,s}^H{\bf{R}}_{h}(\beta){\bf{U}}_{\tau,s}||_F^2}{||{\bf{U}}_{\tau,s}^H{\bf{R}}_{h}(0){\bf{U}}_{\tau,s}-\sigma_n^2{\bf{I}}_L||_F^2}}=\frac{\xi_{\beta}}{\xi_0}
\end{equation}

Through some manipulations (see Appendix \ref{app:xi0andxibeta}),
$\xi_0$ and $\xi_{\beta}$ can be expanded into series and
approximately expressed as
\begin{eqnarray}
\label{eqnarray:xi0expand}
\xi_0&=&\sum\limits_{k=0}^{\infty}s_k\approx\sum\limits_{k=0}^{\infty}\frac{(-{\psi}^2)^k}{k!(k+1)!(2k+1)}\\
\xi_{\beta}&=&\sum\limits_{k=0}^{\infty}t_k\approx\sum\limits_{k=0}^{\infty}\frac{(-{\psi}^2)^k}{k!(k+1)!(2k+1)}\times\nonumber\\
\label{eqnarray:xibetaexpand}
{}&{}&{\;\;\;}{\frac{1}{2}}\left[(1+{\varphi})^{2k+2}+(1-{\varphi})^{2k+2}-2{\varphi}^{2k+2}\right]
\end{eqnarray}
where $N\gg1$, $\psi=\pi{f_d}NT$, and $\varphi=\beta(1+r_{cp})$.

Since the absolute values of $s_k$ and $t_k$ decrease very fast, a
finite number, say, $K$,  of terms in (\ref{eqnarray:xi0expand}) and
(\ref{eqnarray:xibetaexpand}) are sufficient to meet the accuracy
requirement. So, with
(\ref{eqn:etadef})--(\ref{eqnarray:xibetaexpand}), we have
\begin{equation}
\label{eqn:etaexpand} \sum\nolimits_{k=0}^{K-1}(t_k-{\eta}s_k)=0
\end{equation}
Let $x=-{\psi}^2$ and
$c_k=\frac{[(1+{\varphi})^{2k+2}+(1-{\varphi})^{2k+2}-2{\varphi}^{2k+2}]-2\eta}{2k!(k+1)!(2k+1)}$.
Then, (\ref{eqn:etaexpand}) is equivalent to
\begin{equation}
\label{eqn:function} \sum\nolimits_{k=0}^{K-1}c_kx^k=0
\end{equation}

(\ref{eqn:function}) is a non-linear high-order polynomial equation.
By resorting to Newton's method, its root, denoted as $x^*$, is
readily solved after numbers of iterations. Accordingly, $f_d$ is
given by
\begin{equation}
\label{eqn:fdsovle} f_d=\frac{\sqrt{-x^*}}{{\pi}NT}
\end{equation}

When the transceiver is moving, the path delays of the channel are
slowly drifting \cite{Tse05}\cite{Sime04}, which causes
${\bf{F}}_{p;\tau}$ to vary and so does ${\bf{U}}_{\tau,s}$. To
accommodate the variation, the subspace tracking algorithm is
adopted to automatically update ${\bf{U}}_{\tau,s}$. The proposed
algorithm is summarized in the following. It is worth noting that
its computation complexity depends on the QR-decomposition operation
that is ${\mathcal{O}}(P\times{L^2})$. For sparse multipath
channels, the complexity is moderate because $L$ is quite small.
Besides, the choice of $K$ in (\ref{eqn:function}) can be made
according to the tradeoff between the accuracy and complexity. After
$K$ is chosen, the convergence of Newton's method is quadratic along
the number of iterations. In fact, numerical results show that less
than 4 iterations are sufficient to achieve a precision of
$10^{-4}$. Moreover, for each iteration, the first differential is
readily obtained thanks to the polynomial coefficients, which
reduces the complexity of Newton's method considerably. Finally, as
$\xi_{\beta}$ is oscillatingly attenuating along $\beta$,
(\ref{eqn:function}) has multiple roots for large $\beta$
\footnote{For applied OFDM systems, (\ref{eqn:function}) has only
one negative root close to zero when $\beta\leq4$.}. In order to
converge to the specific root, the initialization of Newton's
iteration, therefore, should be carefully chosen according to
$\beta$.

\vspace{-6pt}
\begin{algorithm}
\vspace{2pt}
\begin{algorithmic}
\STATE{\tt\small\textbf{Initialize}: ($n=0$) \vspace{1pt}\\
\begin{array}{lll}
{}&{}&{\bf{Q}}_{0}(0)={\bf{Q}}_{\beta}(0)=[{\bf{I}}_{L_m},{\bf{0}}_{L_m,P-L_m}^T]^T\nonumber\\
{}&{}&{\bf{A}}_{0}(0)={\bf{A}}_{\beta}(0)={\bf{0}}_{P,L_m},\;\;{\bf{C}}_{0}(0)={\bf{C}}_{\beta}(0)={\bf{I}}_{L_m}\nonumber
\end{array} \vspace{1pt}\\
\vspace{1pt}}
\STATE{\tt\small\textbf{Run}: ($n=n+1$)\vspace{1pt}\\
\parindent 1mm Input: ${\bf{h}}_{p,ls}(n)$\vspace{1pt}\\
\parindent 1mm Step 1: Updating eigenvalues of ${\bf{R}}_h(0)$:\vspace{1pt}\\
\begin{array}{rcl}
{\bf{Z}}_{0}(n)&=&{\bf{h}}_{p,ls}(n){\bf{h}}_{p,ls}(n)^H\nonumber\\
{\bf{A}}_{0}(n)&=&\alpha{\bf{A}}_{0}(n-1){\bf{C}}_{0}(n-1)+(1-\alpha){\bf{Z}}_{0}(n){\bf{Q}}_{0}(n-1)^H\nonumber\\
{\bf{A}}_{0}(n)&=&{\bf{Q}}_{0}(n){\bf{R}}_{0}(n) : \textit{\textbf{QR-factorization}}\nonumber\\
{\bf{C}}_{0}(n)&=&{\bf{Q}}_{0}(n-1)^H{\bf{Q}}_{0}(n)\nonumber\\
\end{array} \vspace{0pt}\\
\parindent 1mm Step 2: Updating eigenvalues of ${\bf{R}}_h(\beta)$:\vspace{1pt}\\
\begin{array}{rcl}
{\bf{Z}}_{\beta}(n)&=&{\bf{h}}_{p,ls}(n){\bf{h}}_{p,ls}(n-\beta)^H\nonumber\\
{\bf{A}}_{\beta}(n)&=&\alpha{\bf{A}}_{\beta}(n-1){\bf{C}}_{\beta}(n-1)+(1-\alpha){\bf{Z}}_{\beta}(n){\bf{Q}}_{\beta}(n-1)^H\nonumber\\
{\bf{A}}_{\beta}(n)&=&{\bf{Q}}_{\beta}(n){\bf{R}}_{\beta}(n) : \textit{\textbf{QR-factorization}}\nonumber\\
{\bf{C}}_{\beta}(n)&=&{\bf{Q}}_{\beta}(n-1)^H{\bf{Q}}_{\beta}(n)\nonumber\\
\end{array} \vspace{0pt}\\
\parindent 1mm Step 3: Estimating $L$ and $\sigma_n^2$:\vspace{4pt}\\
\begin{array}{lr}
{\hat{L}}={MDL\left(diag({\bf{R}}_{0}(n))\right)},&{{\hat{\sigma}}_n^2}={\frac{1}{P-{\hat{L}}}\sum\nolimits_{p={\hat{L}}+1}^{P}\left[{\bf{R}}_{0}(n)\right]_{p,p}}\nonumber\\
\end{array} \vspace{4pt}\\
\parindent 1mm Step 4: Estimating $\eta$ according to (\ref{eqn:etadef}):\vspace{4pt}\\
\begin{array}{c}
\hspace{40pt}\hat{\eta}=\sqrt{\frac{\sum\nolimits_{l=1}^{{\hat{L}}(n)}\left|\left[{\bf{R}}_{\beta}(n)\right]_{l,l}\right|^2}{\sum\nolimits_{l=1}^{{\hat{L}}(n)}\left|\left[{\bf{R}}_{0}(n)\right]_{l,l}-{{\hat{\sigma}}_n^2}\right|^2}}\nonumber\\
\end{array} \vspace{4pt}\\
\parindent 1mm Step 5: Estimating $f_d$ by (\ref{eqn:function}) and (\ref{eqn:fdsovle}).\vspace{1pt}\\
}\STATE{\tt\small\textbf{Remark}: $\alpha$ is an exponential
forgetting factor close to 1. $L_{m}$ is the maximum rank to be
tested. $MDL(\cdot)$ denotes the MDL detector. In simulations, we
set $\alpha=0.995$, $\beta=1$, $L_m=10$, $K=8$, and the precision
threshold of Newton iteration and the maximum number of iterations
as $10^{-4}$ and 4, respectively.}
\end{algorithmic}
\end{algorithm}
\vspace{-16pt}

\section{Simulation Results}
\label{sec:simulations} The performance of the proposed algorithm is
evaluated for an OFDM system with $BW=12$ MHz ($T=1/BW=83.3$ ns),
$N=1024$, $L_{cp}=128$ and $P=128$. Two 3GPP E-UTRA channel models
are adopted: Extended Vehicular A model (EVA) and Extended Typical
Urban model (ETU) \cite{3GPP36101}. The delay profile of EVA is
[$0$, $30$, $150$, $310$, $370$, $710$, $1090$, $1730$, $2510$] ns,
and its power profile is [$0.0$, $-1.5$, $-1.4$, $-3.6$, $-0.6$,
$-9.1$, $-7.0$, $-12.0$, $-16.9$] dB. For ETU, they are [$0$, $50$,
$120$, $200$, $230$, $500$, $1600$, $2300$, $5000$] ns and [$-1.0$,
$-1.0$, $-1.0$, $0.0$, $0.0$, $0.0$, $-3.0$, $-5.0$, $-7.0$] dB,
respectively. The classic Doppler spectrum, i.e., Jakes' spectrum
\cite{Stee92}, is applied to generate the Rayleigh fading channels.

\begin{figure}[!t]
 \centering
\includegraphics[width=3.4in,height=2.45in]{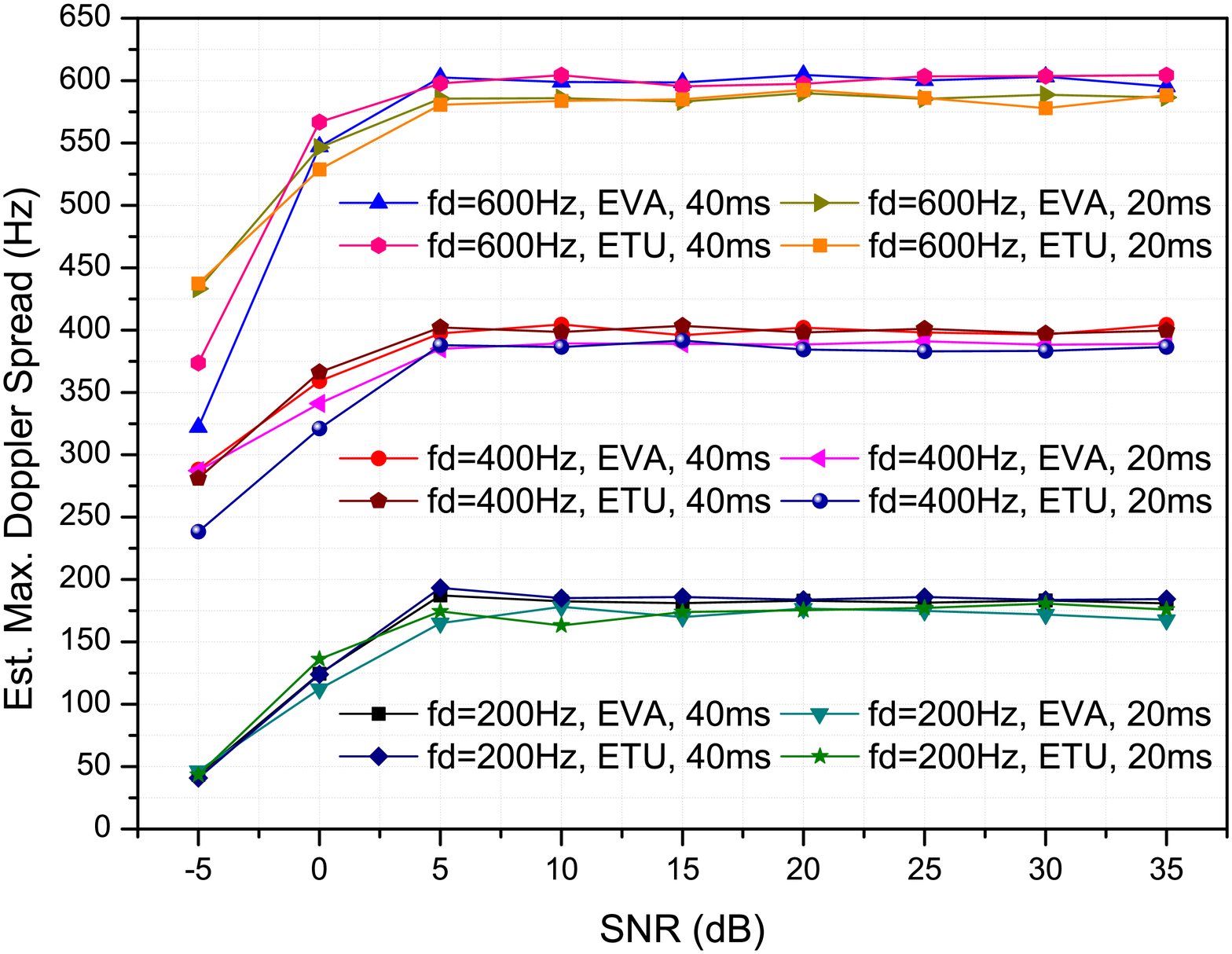}
 \caption{The performance evaluation for the proposed
algorithms for EVA and ETU channels with different amounts of
samples and Doppler's.} \label{fig:fig1} \vspace{-10pt}
\end{figure}

In Fig.\ref{fig:fig1}, the proposed algorithm is evaluated for
$f_d=200$, $400$, and $600$ Hz, respectively, within a wide range of
SNR's. And two durations of observation ($20$ and $40$ ms) are
adopted to acquire the sample auto-correlation matrices. From the
figure, it is evident that the performance of the proposed algorithm
is robust when $SNR\;{\ge}\;5$ dB, since the delay-subspace
effectively reduces the noise disturbance. Furthermore, the
estimation accuracy is rather high for $f_d\;{\ge}\;400$ Hz and the
$40$ ms observation but somewhat deteriorated for $200$ Hz and $20$
ms. It is accounted for the accumulation process of the sample
matrices: the larger the maximum Doppler spread is, the faster the
CFR updates; the longer the duration is, the more sufficient the
sample matrices are. When $f_d$ is yet smaller, say, $50$ Hz, the
proposed algorithm, nevertheless, may fail with the $40$ ms
observation due to the insufficient sample matrices. Besides, a
simplified version of the proposed algorithm \cite{ZhaoGC08} still
outperforms the CP-based algorithm \cite{Cai03}.

\begin{figure}[!t]
 \centering
\includegraphics[width=3.4in,height=2.45in]{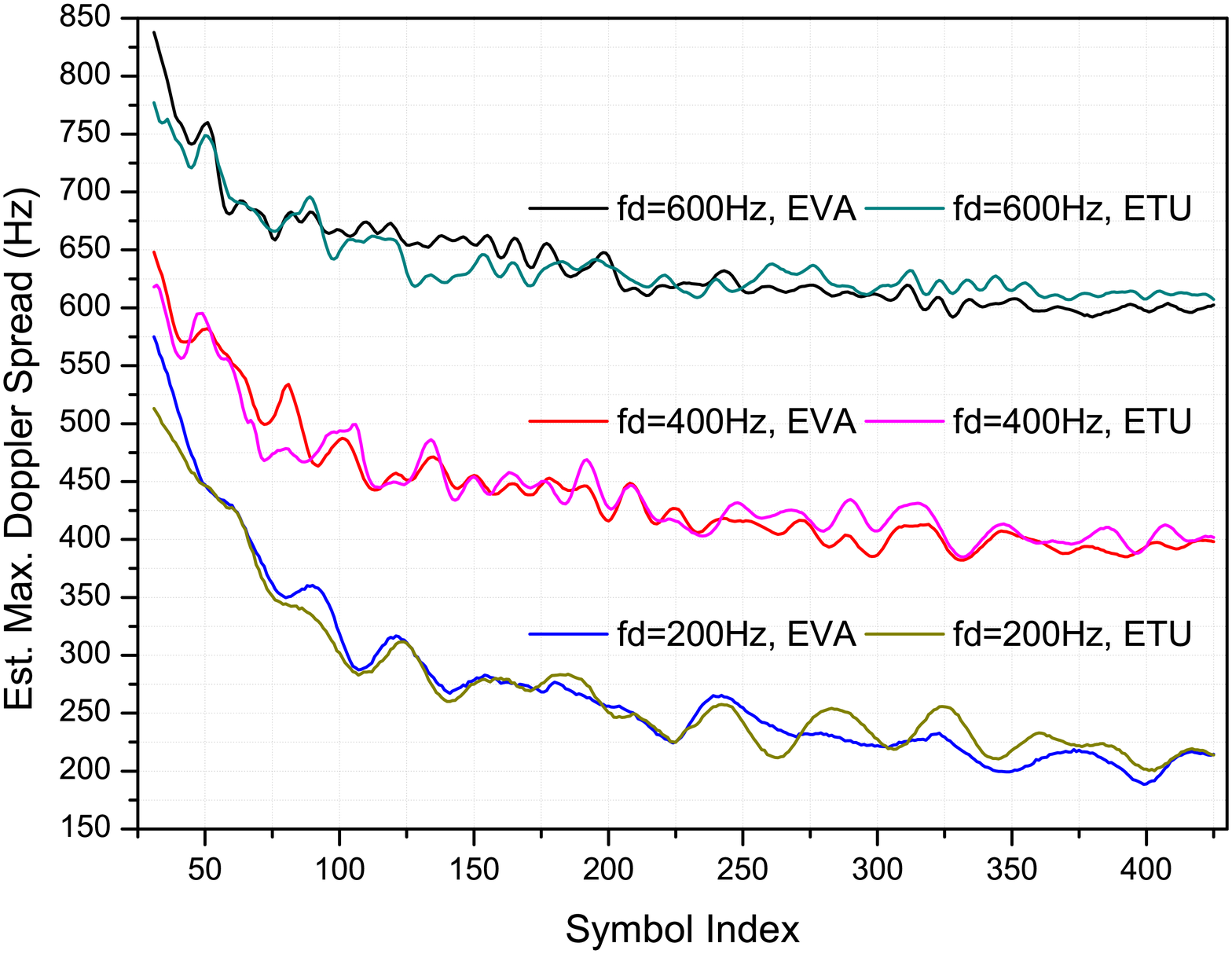}
 \caption{The convergence of the proposed algorithm
when $SNR=15$ dB and $\beta=1$ for EVA and ETU channels.}
\label{fig:fig2} \vspace{-10pt}
\end{figure}

The convergence of the proposed algorithm is verified with various
Doppler's, i.e., $f_d=600$, $400$, $200$ Hz, when $SNR=15$ dB for
EVA and ETU channels, respectively. As plotted in
Fig.\ref{fig:fig2}, the estimated $f_d$'s converge after hundreds of
samples. Furthermore, since the CFR updates faster when $f_d$ is
larger, the convergence speed is faster for larger $f_d$ than
smaller. In addition, for all the curves drawn in
Fig.\ref{fig:fig2}, the estimated values fluctuate around their true
ones within a certain range, and the variations are larger for
smaller $f_d$'s because of the sensitivity to the estimation error
of $\eta$ in (\ref{eqn:etadef}) when $f_dT_s$ is small. If
necessary, an averaging/smoothing window over the output of the
proposed algorithm can be applied to supply a more stable
estimation.

\section{Conclusions}
\label{sec:conclusion} The maximum Doppler spread is crucial for
adaptive strategies in OFDM systems. In this paper, we propose a
subspace-based maximum Doppler spread estimation algorithm
applicable to the comb-type pilot pattern. By tracking the drifting
delay-subspace, the noise is greatly reduced. And by solving the
polynomial equation with the Newton's method, high accuracy can be
achieved with moderate complexity. The performance of the proposed
algorithm is demonstrated to be robust over a wide range of SNR's
and Doppler's by simulations. Besides, the proposed algorithm can be
readily integrated into the channel estimators with the subspace
tracker \cite{Sime04}\cite{Ragh07}, which lends a broad application
promise to it.

\appendices
\section{Approximations of $\xi_0$ and $\xi_{\beta}$}
\label{app:xi0andxibeta} The Maclakutin series of $J_0(z)$ is
\cite{Weis}
\begin{equation}
\label{eqn:J0Macseries}
J_0(z)=\sum\limits_{k=0}^{\infty}\frac{(-1)^k}{2^{2k}(k!)^2}z^{2k}
\end{equation}
hence (\ref{eqn:xibetadef}) is rewritten as
\begin{eqnarray}
\xi_{\beta}&=&\sum\limits_{k=0}^{\infty}t_k=\sum\limits_{k=0}^{\infty}\frac{(-1)^k}{(k!N)^2}\sum\limits_{m_1=0}^{N-1}\sum\limits_{m_2=0}^{N-1}\nonumber\\
\label{eqnarray:tk}
{}&{}&{\;\;\;}\left\{{\pi}f_dT\left[(m_1-m_2)+{\beta}(1+r_{cp})N\right]\right\}^{2k}
\end{eqnarray}
Denoting $\psi=\pi{f_d}NT$ and $\varphi=\beta(1+r_{cp})$, $t_k$ is
further expanded as
\begin{eqnarray}
t_k&=&\frac{(-{\psi}^2)^k}{(k!N)^2}\sum\limits_{m_1=0}^{N-1}\sum\limits_{m_2=0}^{N-1}\sum\limits_{p=0}^{2k}C_{2k}^{p}{\varphi}^{2k-p}\times\nonumber\\
{}&{}&{\;\;\;\;\;}\sum\limits_{q=0}^{p}C_{p}^{q}(\frac{m_1}{N})^{q}(-\frac{m_2}{N})^{p-q}\nonumber\\
{}&=&\frac{(-{\psi}^2)^k}{(k!)^2}\sum\limits_{p=0}^{2k}C_{2k}^{p}{\varphi}^{2k-p}\sum\limits_{q=0}^{p}C_{p}^{q}\times\nonumber\\
\label{eqnarray:tkext}
{}&{}&{\;\;\;\;\;}\left[\frac{1}{N}\sum\limits_{m_1=0}^{N-1}(\frac{m_1}{N})^{q}\right]\left[\frac{1}{N}\sum\limits_{m_2=0}^{N-1}(-\frac{m_2}{N})^{p-q}\right]
\end{eqnarray}
When $N\gg1$, we have
\begin{equation}
\frac{1}{N}\sum\limits_{m_1=0}^{N-1}(\frac{m_1}{N})^{q}\approx\frac{1}{q+1},\;\;\frac{1}{N}\sum\limits_{m_2=0}^{N-1}(-\frac{m_2}{N})^{p-q}\approx\frac{(-1)^{p-q}}{p-q+1}\nonumber
\end{equation}
Thus, $t_k$ can be approximated as
\begin{eqnarray}
t_k&\approx&\frac{(-{\psi}^2)^k}{(k!)^2}\sum\limits_{p=0}^{2k}C_{2k}^{p}{\varphi}^{2k-p}\frac{1+(-1)^p}{(p+1)(p+2)}\nonumber\\
{}&=&\frac{(-{\psi}^2)^k}{k!(k+1)!(2k+1)}\times\nonumber\\
\label{eqnarray:tknew}
{}&{}&{\;\;\;\;}{\frac{1}{2}}\left[(1+{\varphi})^{2k+2}+(1-{\varphi})^{2k+2}-2{\varphi}^{2k+2}\right]
\end{eqnarray}
Similarly, by letting $\beta=0$ in (\ref{eqnarray:tknew}),
(\ref{eqn:xi0def}) is expanded and approximated as
\begin{equation}
\xi_0=\sum\limits_{k=0}^{\infty}s_k\approx\sum\limits_{k=0}^{\infty}\frac{(-{\psi}^2)^k}{k!(k+1)!(2k+1)}
\end{equation}

\bibliographystyle{IEEEtran}
\bibliography{IEEEabrv,SPL_DopplerEst_short}

\begin{thebibliography}{10}
\providecommand{\url}[1]{#1}
\csname url@samestyle\endcsname
\providecommand{\newblock}{\relax}
\providecommand{\bibinfo}[2]{#2}
\providecommand{\BIBentrySTDinterwordspacing}{\spaceskip=0pt\relax}
\providecommand{\BIBentryALTinterwordstretchfactor}{4}
\providecommand{\BIBentryALTinterwordspacing}{\spaceskip=\fontdimen2\font plus
\BIBentryALTinterwordstretchfactor\fontdimen3\font minus
  \fontdimen4\font\relax}
\providecommand{\BIBforeignlanguage}[2]{{%
\expandafter\ifx\csname l@#1\endcsname\relax
\typeout{** WARNING: IEEEtran.bst: No hyphenation pattern has been}%
\typeout{** loaded for the language `#1'. Using the pattern for}%
\typeout{** the default language instead.}%
\else
\language=\csname l@#1\endcsname
\fi
#2}}
\providecommand{\BIBdecl}{\relax}
\BIBdecl

\bibitem{Chu00}
M.~Chu \emph{et~al.}, ``{Effect of Mobile Velocity on Communication in Fading
  Channels},'' \emph{{IEEE} Trans. Veh. Technol.}, vol.~49, pp. 202--210, Jan.
  2000.

\bibitem{Tepe01}
C.~Tepedelenlioglu \emph{et~al.}, ``{E}stimation of {D}oppler {S}pread and
  {S}ignal {S}trength in {M}obile {C}ommunications with {A}pplications to
  {H}andoff and {A}daptive {Transmission},'' \emph{Wirel. Commun. Mob.
  Comput.}, vol.~1, pp. 221--242, Aug. 2001.

\bibitem{Tepe01VT}
C.Tepedelenlioglu \emph{et~al.}, ``{On Velocity Estimation and Correlation
  Properties of Narrow-Band Mobile Communication Channels},'' \emph{{IEEE}
  Trans. Veh. Technol.}, vol.~50, pp. 1039--1052, Jul. 2001.

\bibitem{Scho02}
H.~Schober \emph{et~al.}, ``{V}elocity {E}stimation for {OFDM} {B}ased
  {C}ommunication {S}ystems,'' in \emph{IEEE VTC-Fall 2002}, vol.~2, Vancouver,
  BC, Canada, 2002, pp. 715--718.

\bibitem{Cai03}
J.~Cai \emph{et~al.}, ``{D}oppler {S}pread {E}stimation for {M}obile {OFDM}
  {S}ystems in {R}ayleigh {F}ading {C}hannels,'' \emph{{IEEE} Trans. Consum.
  Electron.}, vol.~49, pp. 973--977, Nov. 2003.

\bibitem{Yuce05}
T.~Yucek \emph{et~al.}, ``{D}oppler {S}pread {E}stimation for {W}ireless {OFDM}
  {S}ystems,'' in \emph{IEEE/Sarnoff Symposium on Advances in Wired and
  Wireless Communication}, 2005, pp. 233--236.

\bibitem{Cole02}
S.~Coleri \emph{et~al.}, ``{C}hannel {E}stimation {T}echniques {B}ased on
  {P}ilot {A}rrangement in {OFDM} {S}ystems,'' vol.~48, pp. 223--229, Sep.
  2002.

\bibitem{Sime04}
O.~Simeone \emph{et~al.}, ``{P}ilot-{B}ased {C}hannel {E}stimation for {OFDM}
  {S}ystems by {T}racking the {D}elay-{S}ubspace,'' vol.~3, pp. 315--325, Jan.
  2004.

\bibitem{Stro96}
P.~Strobach, ``{L}ow-{R}ank {A}daptive {F}ilters,'' \emph{{IEEE} Trans. Signal
  Process.}, vol.~44, pp. 2932--2947, Dec. 1996.

\bibitem{Ohno04}
S.~Ohno \emph{et~al.}, ``{C}apacity {M}aximizing {MMSE}-{O}ptimal {P}ilots for
  {W}ireless {OFDM} {O}ver {F}requency-{S}elective {B}lock {R}ayleigh-{F}ading
  {C}hannels,'' \emph{{IEEE} Trans. Inf. Theory}, vol.~50, pp. 2138--2145, Sep.
  2004.

\bibitem{Stee92}
R.~Steele, \emph{{M}obile {R}adio {C}ommunications}.\hskip 1em plus 0.5em minus
  0.4em\relax IEEE Press, 1992.

\bibitem{Clar68}
R.~Clarke, ``{A} {S}tatistical {T}heory of {M}obile {R}adio {R}eception,''
  \emph{Bell Syst. Tech. J.}, pp. 957--1000, Jul.-Aug. 1968.

\bibitem{Choi01}
Y.~Choi \emph{et~al.}, ``{O}n {C}hannel {E}stimation and {D}etection for
  {M}ulticarrier {S}ignals in {F}ast and {S}elective {R}ayleigh {F}ading
  {C}hannels,'' \emph{{IEEE} Trans. Commun.}, vol.~49, pp. 1375--1387, Aug.
  2001.

\bibitem{Wax85}
M.~Wax \emph{et~al.}, ``{D}etection of {S}ignals by {I}nformation {T}heoretic
  {C}riteria,'' \emph{{IEEE} Trans. Acoust., Speech, Signal Process.}, vol.~33,
  pp. 387--392, Apr. 1985.

\bibitem{Tse05}
D.~Tse \emph{et~al.}, \emph{{F}undamentals of {W}ireless
  {C}ommunication}.\hskip 1em plus 0.5em minus 0.4em\relax New York: Cambridge
  University Press, 2005.

\bibitem{3GPP36101}
``{3GPP} {TS} 36.101 v8.2.0 -- {E}volved {U}niversal {T}errestrial {R}adio
  {A}ccess ({E-UTRA}); {U}ser {E}quipment ({UE}) {R}adio {T}ransmission and
  {R}eception ({R}elasase 8),'' 3GPP, May 2008.

\bibitem{ZhaoGC08}
{X. Zhao and others}, ``{Doppler Spread Estimation by Subspace Tracking for
  OFDM Systems},'' in \emph{{IEEE GLOBECOM 2008}}, {New Orleans, LA, USA}, Nov.
  {2008}.

\bibitem{Ragh07}
M.~Raghavendra \emph{et~al.}, ``{P}arametric {C}hannel {E}stimation for
  {P}seudo-{R}andom {T}ile-{A}llocation in {U}plink {OFDMA},'' \emph{{IEEE}
  Trans. Signal Process.}, vol.~55, pp. 5370--5381, Nov. 2007.

\bibitem{Weis}
\BIBentryALTinterwordspacing
E.~Weisstein, ``{B}essel {F}unction of the {F}irst {K}ind,'' From MathWorld - A
  Wolfram Web Resource. [Online]. Available:
  \url{http://mathworld.wolfram.com/BesselFunctionoftheFirstKind.html}
\BIBentrySTDinterwordspacing

\end{thebibliography}

\setlength{\arraycolsep}{5pt}

\end{document}